\documentclass{aa}  

\usepackage{graphicx}
\usepackage{txfonts}
\begin{document} 

    \title{CENN: A fully convolutional neural network for CMB recovery in realistic microwave sky simulations}
    \titlerunning{CMB recovery with FCN}
    \authorrunning{Casas J.M. et al.}
   
    \author{Casas J. M.\inst{1,2},
    Bonavera L. \inst{1,2},
    Gonz{\'a}lez-Nuevo J.\inst{1,2},
    Baccigalupi, C.\inst{3,4,5},
    Cueli M. M.\inst{1,2},
    Crespo D. \inst{1,2},
    Goitia E. \inst{1},
    Santos J. D. \inst{1,2},
    S{\'a}nchez M. L. \inst{1,2},
    de Cos F. J. \inst{2,6}
}

   \institute{$^1$Departamento de F{\'i}sica, Universidad de Oviedo, C. Federico Garc{\'i}a Lorca 18, 33007 Oviedo, Spain\\
    \email{casasjm@uniovi.es}\\
             $^2$Instituto Universitario de Ciencias y Tecnolog{\'i}as Espaciales de Asturias (ICTEA), C. Independencia 13, 33004 Oviedo, Spain\\
             $^3$SISSA, Via Bonomea 265, 34136 Trieste, Italy\\
             $^4$IFPU - Institute for fundamental physics of the Universe, Via Beirut 2, 34014 Trieste, Italy\\
             $^5$INFN-Sezione di Trieste, via Valerio 2, 34127 Trieste,  Italy\\
             $^6$Escuela de Ingeniería de Minas, Energía y Materiales Independencia 13, 33004 Oviedo, Spain\\
             }


 \abstract{}{}{}{}{} 
  \abstract
   {Component separation is the process with which emission sources in astrophysical maps are generally extracted by taking multi-frequency information into account. It is crucial to develop more reliable methods for component separation for future cosmic microwave background (CMB) experiments such as the Simons Observatory, the CMB-S4, or the LiteBIRD satellite.}
   {We aim to develop a machine learning method based on fully convolutional neural networks called the cosmic microwave background extraction neural network (CENN) in order to extract the CMB signal in total intensity by training the network with realistic simulations. The frequencies we used are the \textit{Planck} channels 143, 217, and 353 GHz, and we validated the neural network throughout  the sky and at three latitude intervals: $0^\circ<|b|<5^\circ$, $5^\circ<|b|<30^\circ$ , and $30^\circ<|b|<90^\circ$. Moreover, we used neither Galactic nor point-source (PS) masks.}
   {To train the neural network, we produced multi-frequency realistic simulations in the form of patches of 256\hspace{1pt}$\times$\hspace{1pt}256 pixels that contained the CMB signal, the Galactic thermal dust, cosmic infrared background, and PS emissions, the thermal Sunyaev-Zel'dovich effect from galaxy clusters, and instrumental noise. After validating the network, we compared the power spectra from input and output maps. We analysed the power spectrum from the residuals at each latitude interval and throughout the sky, and we studied how our model handled high contamination at small scales.}
   {We obtained a CMB power spectrum with a mean difference between input and output of 13 $\pm$ 113 $\mu K^{2}$ for multipoles up to above 4 000. We computed the residuals, obtaining 700 $\pm$ 60 $\mu K^{2}$ for  $0^\circ<|b|<5^\circ$, 80 $\pm$ 30 $\mu K^{2}$ for $5^\circ<|b|<30^\circ$ , and 30 $\pm$ 20 $\mu K^{2}$ for $30^\circ<|b|<90^\circ$ for multipoles up to above 4 000. For the entire sky, we obtained 30 $\pm$ 10 $\mu K^{2}$ for $l\le1000$ and 20 $\pm$ 10 $\mu K^{2}$ for $l\le 4000$. We validated the neural network in a single patch with strong contamination at small scales, obtaining a difference between input and output of 50 $\pm$ 120 $\mu K^{2}$ and residuals of 40 $\pm$ 10 $\mu K^{2}$ up to $l \sim$ 2 500. In all cases, the uncertainty of each measure was taken as the standard deviation.}
   {The results show that fully convolutional neural networks are promising methods for performing component separation in future CMB experiments. Moreover, we show that CENN is reliable against different levels of contamination from Galactic and PS foregrounds at both large and small scales.}


   \maketitle
%

\section{Introduction}

The cosmic microwave background (CMB) is the relic emission from the primordial Universe at an age of about 380 000 years, an epoch called recombination \citep{SCH06}. At first aproximation, it is homogeneus and isotropic. However, it has small deviations in intensity (temperature) and polarisation from point to point in the sky. These are called anisotropies. They were firstly detected by the Cosmic Background Explorer satellite (COBE; \citealt{SMO92}), and the detection allow fitting the cosmological parameters and cemented the gravitational instability paradigm within a cold dark matter model. After this, the \textit{Wilkinson} Microwave Anisotropy Probe (WMAP; \citealt{BEN13}) showed that the fluctuations are predominantly adiabatic. Finally, \textit{Planck} \citep{PLA_18_I} obtained a more precise CMB signal for both intensity and polarisation with its angular resolution, which is higher than that the previous instruments.

The CMB anisotropies are one of the most important fields of research in modern cosmology. They are divided into two types: primary anisotropies, which are temperature fluctuations that originated in the recombination era, and secondary anisotropies, which are due to distortions of photons through their propagation along the Universe. These anisotropies can be studied by estimating the CMB power spectra, which describes the amplitude of the fluctuations $l \hspace{1pt} (l + 1) \hspace{1pt} C_{l}$ on an angular scale $\theta \sim \pi/l = 180^\circ/l$.

Before recombination, photons and baryons are a tightly coupled photon-baryon fluid \citep{PEE70} that oscillates relativistically, causing a first peak at $l_{1} \sim 200$ and additional maxima at approximately integer multiples of $l_{1}$, called acoustic peaks. These peaks provide fundamental information about cosmology and, in particular, about the initial conditions and the energy contents of the Universe. Therefore, it is very important to measure them with the highest precision, implying a very precise recovery of the CMB signal. However, CMB measurements are contaminated by other signals called foregrounds (\citealt{KRA16}, \citeyear{KRA18}). 

The process of cleaning the microwave maps in order to recover the CMB signal is called component separation \citep{BAC03}. It usually exploits multi-frequency sky maps to extract cosmological information. Our work focuses on \textit{Planck} central wavelengths, where the foregrounds are divided into two categories: firstly, diffuse emission from our Galaxy, which is mostly dominated by thermal dust emission \citep{HEN21}. This is the main contaminant at large angular scales. Secondly, extragalactic foregrounds such as proto-spheroidal galaxies forming the cosmic infrared background (CIB; \citealt{Dol06}), radio galaxies \citep{deZOT09}, infrared (IR) dusty galaxies in the process of forming their stellar masses \citep{Tof98}, and also both thermal and kinetic Sunyaev-Zeldovich effects from galaxy clusters \citep{car02}.

Component separation was one of the central objectives of the \textit{Planck} mission (\citealt{LEA08}, \citeyear{PLA_15_IX}, \citeyear{PLA_15_X}). The mission used four different methods. They were chosen because each of the selected method relies on a different class of algorithms (i.e. blind and non-blind methods, pixel-based, and harmonic-based methods). The first algorithm was NILC \citep{del09}, which makes minimum assumptions for the foregrounds in order to minimise the variance of the CMB using a wavelet on the sphere on the internal linear combination (ILC) algorithm. The second algorithm was SEVEM \citep{MartinezGonzalez2003}, which is a template-based method that removes foreground contamination  from the CMB-dominant bands from the lowest and highest frequency channels. The third algorithm was based on a Bayesian parameter estimation technique called Commander \citep{ERI06}, and the fourth method was SMICA \citep{DEL03}, a parametric approach that allows fitting the amplitude and spectral parameters of the CMB and foregrounds. The main issue of the four methods is that they perform by masking both the Galactic plane and the point sources in order to avoid their contamination. Their final confidence mask in total intensity keeps 78\% of the sky. Moreover, the implementation of these methods required years of research and developing in order to reach the current precise CMB measurement, with the introduction of notable improvements from the early to latest releases. It is the not always straightforward application of traditional methods that drives our interest in trying a different approach to component separation. 

The amount of data in the past years has been produced a high impact in fields such as astrophysics and cosmology. It is necessary to develop automatic and reliable methods that can perform on the new data. The most promising methods are those based on machine learning (ML) because once they are trained, they can be validated on new data, providing results in a much shorter period of computational time with respect to the whole process of the traditional approaches. Some of the most important ML methods are artificial neural networks, mathematical approaches inspired by neuroscience. They can optimise models with non-linear behaviours. Some recent applications in cosmology were in the PS detection field, both in single-frequency \citep{BON21} and in multi-frequency \citep{CAS22} approaches. Moreover, they have been used in the statistical study of the CMB for extending foreground models to sub-degree angular scales \citep{KRA21} and to perform a foreground model-recognition for B-mode observations \citep{FAR20}. They have also been used in works similar to ours for CMB recovery and thermal dust cleaning (\citealt{PET20};  \citealt{WAN22}; an all sky approximation for multipoles up to l $\sim$ 900 in the first case, and in 2D patches for multipoles up to l $\sim$ 1500 in the second). Another timely application is the model by \citet{JEF22}, a moment network for polarised foreground removal using a single training image.

The aim of this work is to train a fully convolutional neural network, called the cosmic microwave background extraction neural network (CENN), for the recovery of the CMB signal with realistic microwave sky simulations in total intensity to determine the potential of the method. We study polarisation data in an upcoming work. The typical application of fully convolutional networks (FCN) is in the Euclidean space, that is, to flat images. This clearly allows a CMB power spectrum estimation only down to angular scales allowed by the patch dimension. This approach might be directly applied to ground-based experiments that will observe limited regions of the sky \citep[e.g. the Simons Observatory and CMB-S4;][]{SO, CMBS4}.
In the future, for full-sky coverage experiments \citep[e.g. the LiteBIRD satellite;][]{SUG20}, it would be useful to apply FCN to the Hierarchical Equal Area Latitude Pixelization scheme (HEALPix; \citealt{GOR05}) using for example the approach by \cite{KRA19}.

The outline of this paper is the following. Section \ref{sec:simulations} covers the generation of the simulated maps. Section \ref{sec:methodology} describes our method. The results are explained in Section \ref{sec:results}, and our conclusions are discussed in Section \ref{sec:conclusions}.
\section{Simulations}
\label{sec:simulations}

\begin{figure*}[ht]
\centering
\includegraphics[width=15.5cm, height=3.5cm]{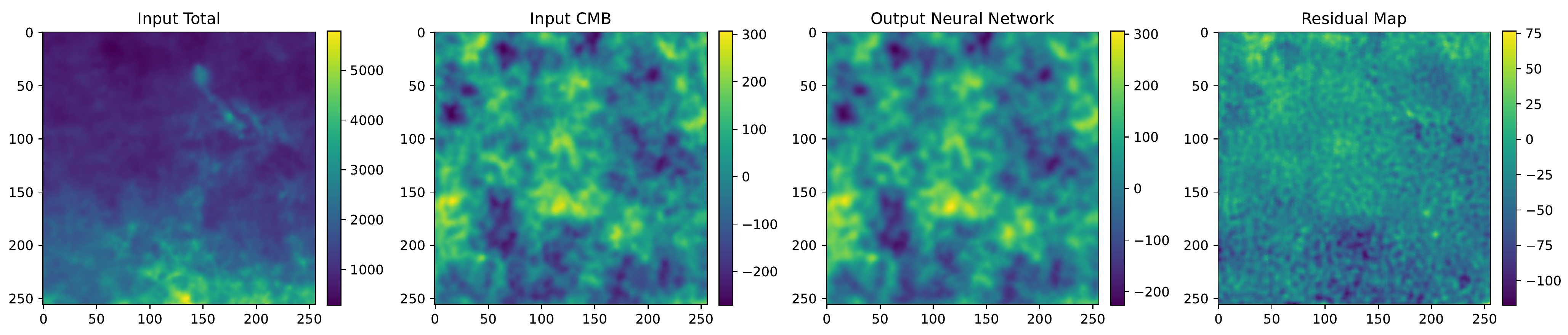}
\includegraphics[width=15.5cm, height=3.5cm]{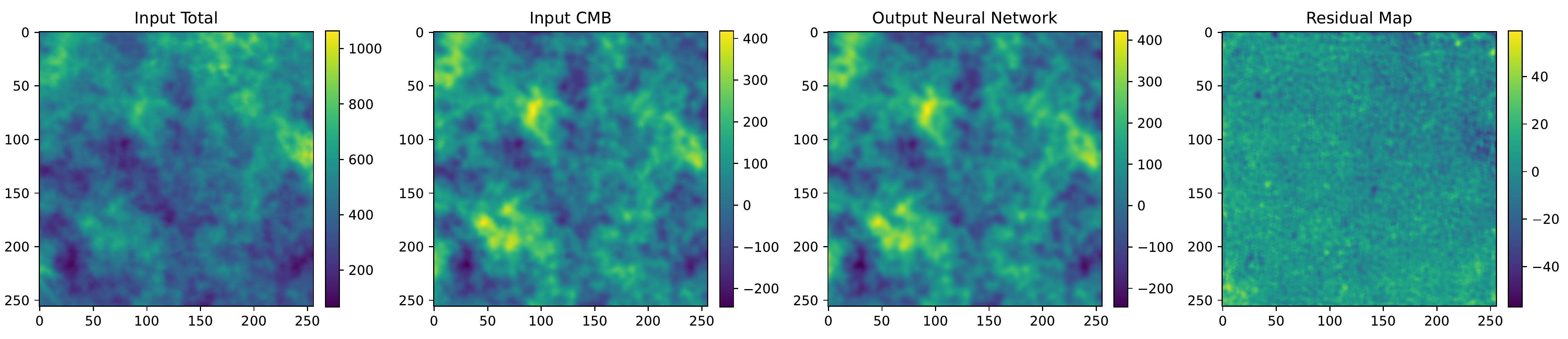}
\includegraphics[width=15.5cm, height=3.5cm]{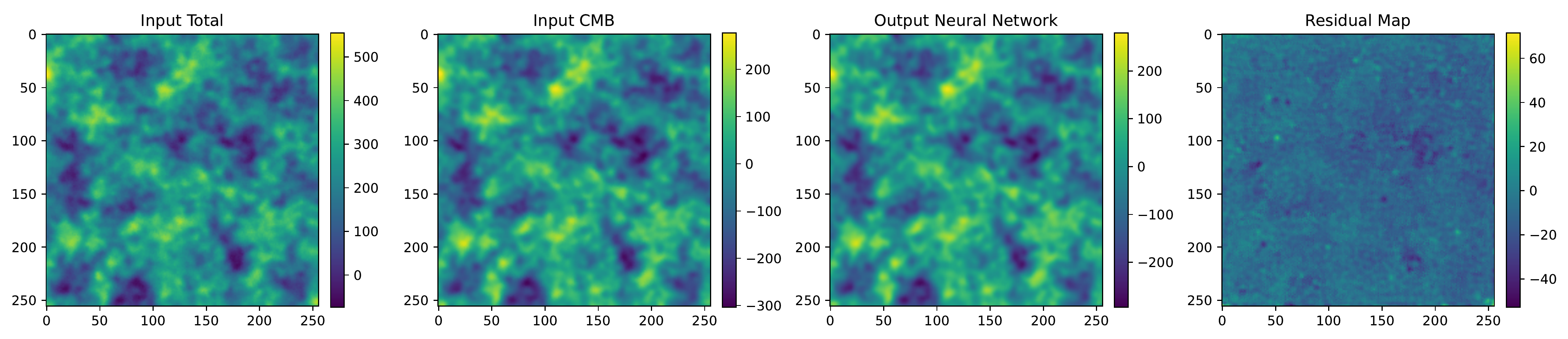}
\caption{Patches at $0^\circ<|b|<5^\circ$ (top row), $5^\circ<|b|<30^\circ$ (middle row), and $30^\circ<|b|<90^\circ$ (bottom row) latitude intervals. They represent the maps with all the emissions, the CMB label, the neural network output, and the residual map computed as the difference between the second and the third column patches from left to right. The frequency is 217 GHz, and their temperature (in $\mu K_{CMB}$) is shown in the right vertical bars. The \textit{x} and \textit{y} labels are the pixel coordinates for each 256\hspace{1pt}$\times$\hspace{1pt}256 patch.}
\label{Fig 1.}
\end{figure*}

To train CENN, we used realistic simulations of the microwave sky. Each simulation was formed by sky patches at the central channels of the \textit{Planck} mission: 143, 217, and 353 GHz. Each patch has an area of 256\hspace{1pt}$\times$\hspace{1pt}256 pixels in order to have good statistics with the shortest possible computation time. The pixel size was 90 arcsec, a number close to the 1.72 arcmin used in the \textit{Planck} mission, which correspond to $N_{\rm side}=2048$ in the \texttt{HEALPix} all-sky pixelisation schema. The maps were randomly simulated at any position in the sky, without the use of PS or Galactic masks, and they were selected in such a way that the probability of having two exact sky maps was negligible. We simulated different datasets: one formed by 60 000 simulations used for training, another one with 6 000 simulations to evaluate the model during that training, and validation datasets with simulations apart from the train one in order to determine how CENN performs on new data. The first dataset consisted of 6 000 simulations of patches centred at any position in the sky (all-sky case). Other three datasets of 2 000 simulations each were built with patches centred at three different latitude intervals: $0^\circ < |b| < 5^\circ$ (inner Galactic region), $5^\circ < |b| < 30^\circ$ (intermediate Galactic region), and $30^\circ < |b| < 90^\circ$(extragalactic region).

First, the CMB signal was projected from all-sky simulations available in the Planck Legacy Archive (PLA\footnote{http://pla.esac.esa.int/pla/\#home}), which were produced using the Planck Sky Model software (PSM; \citealt{Del13}). The CMB signal used in the train and test datasets was taken from a different PLA simulation than we used in the validation dataset.

As Galactic foregrounds, we assumed that only thermal dust emission contaminate the CMB, and we considered it by adding FFP10 PLA simulations from the third$^{}$ \textit{Planck} release. Although it is a good modelling at first approximation, some studies predict significant variations of the two modified blackbody model of the simulations used in this work (\citealt{BOU05}, \citealt{MEN07}, \citealt{PAR11}), which directly affect the behaviour of dust at small scales, neglecting the real non-Gaussian structures of this foreground. 

As extragalactic foregrounds, firstly, the thermal Sunyaev-Zel'dovich effect produced by galaxy clusters was also taken into account using FFP10 PLA simulations as well. Furthermore, point sources (PS) were added as two different populations: the first population consisted of radio galaxies, mainly radio quasars and BL Lacertae objects, commonly known as blazars. These were simulated in the central channel (i.e 217 GHz) following the model by \citet{TUC11} with the software \texttt{CORRSKY} \citep{GN05}. The second population contained IR late-type galaxies, mainly starburst and local spiral galaxies (IRLT), which were simulated with \texttt{CORRSKY} in the central channel using the model by \citet{CAI13} and normalised following the update by \citet{Neg13}. Their spectral behaviour was considered following \citet{CAS22} by assuming that their emission at 143 and 353 GHz varied as
\begin{equation}
    S \hspace{2pt} = \hspace{2pt} S_{0} \hspace{1pt} \left( \frac{\nu}{\nu_{0}} \right)^{\alpha},
\label{eq:flux_density}
\end{equation}
where $S$ is the flux density of the sources at each frequency $\nu$, $S_{0}$ is the flux density in the central channel with frequency $\nu_{0}$, and $\alpha$ is the spectral index, which was randomly selected for each population using the distributions given in \citealt{PCCS2}. 

Proto-spheroidal galaxies were also considered as the elements forming the cosmic IR background (CIB; \citealt{Hau01}), a main contaminant at high frequencies. They were simulated using the source number counts given by \citealt{CAI13}, the angular power spectrum obtained by \citealt{LAP11} , and the software \texttt{CORRSKY}. Their spectral behaviour was also considered following equation \eqref{eq:flux_density}, assuming that they have the same spectral index distribution as the IRLT population. 

Finally, instrumental noise was added to the total input maps in order to simulate the \textit{Planck} HFI instrument. To do this, we used 0.55, 0.78, and 2.56 $\mu K_{CMB}$ deg values for 143, 217, and 353 GHz, respectively \citep{PLA18_IV}.

Each simulation was formed by two different sets of maps called the input total images and the labels. The input total images are multi-frequency realistic simulations of the microwave sky at 143, 217, and 353 GHz formed by the CMB added to other signals acting as contaminants. The labels are maps with only the CMB signal at 217 GHz. To project the maps into flat patches, the \texttt{gnomview} function of \texttt{HEALPix} framework was used. An example of simulations at different latitudes is shown in the first two columns of Figure \ref{Fig 1.} for 217 GHz. The first column represents the total input images with both the CMB signal and the other contaminants, and the second column shows the CMB label maps.


\section{Method}
\label{sec:methodology}

ML is the ability of a computer program to learn some class of tasks and performance measures from experience. Neural networks are one of the most popular ML methods. They are inspired by neuroscience and are generally used to learn non-linear parameters in a model. They are usually composed of connected layers with basic computing units called neurons, which adjust their weights and bias values in each training step.

One of the most popular applications of neural networks is to learn patterns in images to perform a detection using a special type of architecture called convolutional neural networks \citep{LeC89}. Image segmentation is an evolution of this task because it consists of classifying each pixel instead of the whole image. This allows separating one class from the others. 
Fully convolutional neural networks (FCN; \citealt{LON15}) are used to perform image segmentation. They learn parameters by extracting the main features of an image through a set of convolutional blocks, while they make predictions through a set of deconvolutional blocks. Each convolutional block is formed by a layer that reads information from a matrix of input data, and after this, it performs convolutions in parallel. A set of activation functions linearize the convolutions, and a pooling layer aggregates information by grouping neighbouring pixels using their local statistics. Deconvolutional blocks have a similar architecture as convolutional blocks, but in this case, the linearized information is deconvolved, resulting in a segmentation of the class that we wish the neural network to learn. 

Like other neural networks, FCN learn by performing optimisation through minimising a loss function located at the end of their architecture, a process called forward propagation. To drive the loss function to a minimum, a gradient-based optimiser is used on a small set of samples from the training dataset. This is called a minibatch. When the gradient is updated, the model parameters, which adjust some components of the architecture known as the filters, the weights, or the bias, are also updated. This process is called back propagation (\citealt{Rum86}). It results in a theoretically better knowledge of the information read in each epoch, which is the interval within which the entire train dataset flows forward and backward in the neural network. 

\subsection{Model}
\label{sec:Model}

On the one hand, our model is based on the physical assumptions explained in \citealt{ERI06}, \citeyear{ERI08}. On the other hand, it follows the mathematical behaviours of CNN and FCN, which are extensively revised in \citealt{Rum86}, \citealt{LeC98} and \citealt{GOO16}. In this subsection, we explain the physical concepts of our model. The mathematical concepts are summarised in Appendix \ref{sec:appendix}.

The model consists of patches $x_{\nu}$ of the microwave sky at frequency $\nu$ (143, 217, and 353 GHz in this work), each one composed of a linear combination of $c$ astrophysical components with ${e_{\nu}}$ emission maps convolved with the instrumental beam of the experiment $i_{\nu}$ plus instrumental noise $n_{\nu}$, that is,
\begin{equation}
    x_{\nu} \hspace{1pt} = \hspace{1pt} i_{\nu} \hspace{1pt} * \hspace{1pt} e_{\nu} \hspace{1pt} + \hspace{1pt} n_{\nu},
\label{eq:model}
\end{equation}
where the asterisk  denotes convolution. Moreover, the emission maps are a superposition of different astrophysical components, each one with a particular non-linear spectral behaviour. Their superposition can be written as a linear combination of each foreground,
\begin{equation}
    e_{\nu} \hspace{1pt} = \hspace{1pt} \sum_{c} \hspace{1pt} A_{c, \hspace{1pt} \nu} \hspace{1pt} s_{c},
\label{eq:emission_maps}
\end{equation}
where $A$ is called the mixing matrix, and $s$ is the vector of components. Therefore, equation \eqref{eq:model} becomes
\begin{equation}
    x_{\nu} \hspace{1pt} = i_{\nu} \hspace{1pt} * \hspace{1pt} \left( \sum_{c} \hspace{1pt} A_{c, \hspace{1pt} \nu} \hspace{1pt} s_{c} \right) \hspace{1pt} + \hspace{1pt} n_{\nu}.
\label{eq:final_model}
\end{equation}
In the first epoch, the parameters of the model are randomly initialised, and the neural network reads a tensor formed by a matrix of $N_{pix} \times N_{pix} \times N_{\nu}$ in its first block, where $N_{pix}=256$ is the number of pixels, and $N_{\nu}$ is the number of frequency channels. Then, the spectral dimension of the input data is considered by estimating a multidimensional stride convolution \citep{GOO10},
\begin{equation}
    H_{\nu, \hspace{1pt} i, \hspace{1pt} j} \hspace{1pt} = \hspace{1pt} \sum_{k, \hspace{1pt} m} W_{k, \hspace{1pt} m, \hspace{1pt} i} \hspace{1pt} \hspace{1pt} V_{s\hspace{1pt} \circ \hspace{1pt} \nu \hspace{1pt} + \hspace{1pt} k, \hspace{1pt} m, \hspace{1pt} j}\hspace{1pt},
\label{eq:convolutional_blocks}
\end{equation}
where $H$ are the hidden units (i.e related to the resulting outputs in the first convolutional block), which index each spectral dimension $\nu$ within the feature map $i$ for sample $j$. $W$ is the kernel, which connects the weights between $i$ and $j$ for each spectral channel.  $V$ are the visible units (i.e related to the input images), which have the same format as $H$. The subindex $s$ represents a vector of strides\footnote{They are the number of pixels shifts over the input matrix. Because in our case, $s=2$ in every layer, the network move the filters to two pixels at a time in each layer.}. The $\text{open circle}$ represents the elementwise product. The next convolutional blocks have similar operations, but they consider the information computed in their previous convolutional block as inputs.

In the deconvolutional blocks, the spectral dimensions of the data are considered by computing a multidimensional transpose of strided convolution
\begin{equation}
    R_{q, \hspace{1pt}m, \hspace{1pt}j} \hspace{1pt}= \sum_{\mathbf{\nu}, \hspace{1pt} k \hspace{1pt}| \hspace{1pt}\mathbf{s} \hspace{1pt}\circ \hspace{1pt}\mathbf{\nu} \hspace{1pt}+\hspace{1pt} k \hspace{1pt}= \hspace{1pt} q} \hspace{1pt} \sum_{i} \hspace{1pt} W_{k, \hspace{1pt}m, \hspace{1pt}i} \hspace{1pt} H_{\nu,\hspace{1pt} i,\hspace{1pt} j} \hspace{1pt},
\label{eq:deconvolutional_blocks}
\end{equation}
where $R$ are the hidden units in the deconvolutional block, and $H$ are the visible units (i.e the information from the previous block), which index each spectral dimension $q$ within the feature map $m$ for the sample $j$. The vertical line  denotes the modulus operator\footnote{It is used to sum over the correct set of values for $\nu$ and $k$.}. The other terms are those defined in Equation \eqref{eq:convolutional_blocks}. The kernel values $W$ in all convolutional and deconvolutional blocks are updated on each epoch after minimising the mean-squared error loss function, 
\begin{equation}
    MSE \hspace{1pt} = \hspace{1pt} \frac{1}{2} \hspace{1pt} | \hspace{1pt} y \hspace{1pt} - \hspace{1pt} y\hspace{1pt}' \hspace{1pt} |\hspace{1pt}^{2},
\label{eq:mse}
\end{equation}
where $y$ is the value predicted by the model, and $y\hspace{1pt}'$ is the true value, which in our case is the map with the true CMB emission. This function is located at the end of the architecture, and after computing it, a gradient is then estimated. When the gradient is computed, the parameters for each layer become updated as explained in Appendix \ref{sec:appendix}. In the last deconvolutional block, all the spectral information of its previous block is deconvolved into a single channel in order to predict the CMB signal at $\nu_{0} =$ 217 GHz, that is, the network computes the quantity $R_{q\hspace{1pt}=\hspace{1pt}\nu_{0}, \hspace{1pt}m, \hspace{1pt}j}$ by indexing $H_{\mathbf{\nu},\hspace{1pt} i,\hspace{1pt} j}$ in Equation \eqref{eq:deconvolutional_blocks} over all the filters on the previous deconvolutional block, producing the quantity
\begin{equation}
    \Tilde{x}_{\nu_{0}} \hspace{1pt} = \hspace{1pt} i_{\nu_{0}} \hspace{1pt} * \hspace{1pt} e_{CMB, \hspace{1pt} \nu_{0}} \hspace{1pt} + \hspace{1pt} \Tilde{n},
\label{eq:final_map}
\end{equation}
that is, we obtain a map $\Tilde{x}_{\nu_{0}}$ at $\nu_{0}$ frequency composed of the $e_{CMB, \hspace{1pt} \nu_{0}}$ CMB emission convolved with $i_{\nu_{0}}$ (i.e the instrumental beam of the experiment) plus $\Tilde{n}$, which is generalised noise due to deconvolution.

\begin{figure*}[ht]
\centering
\includegraphics[width=17cm]{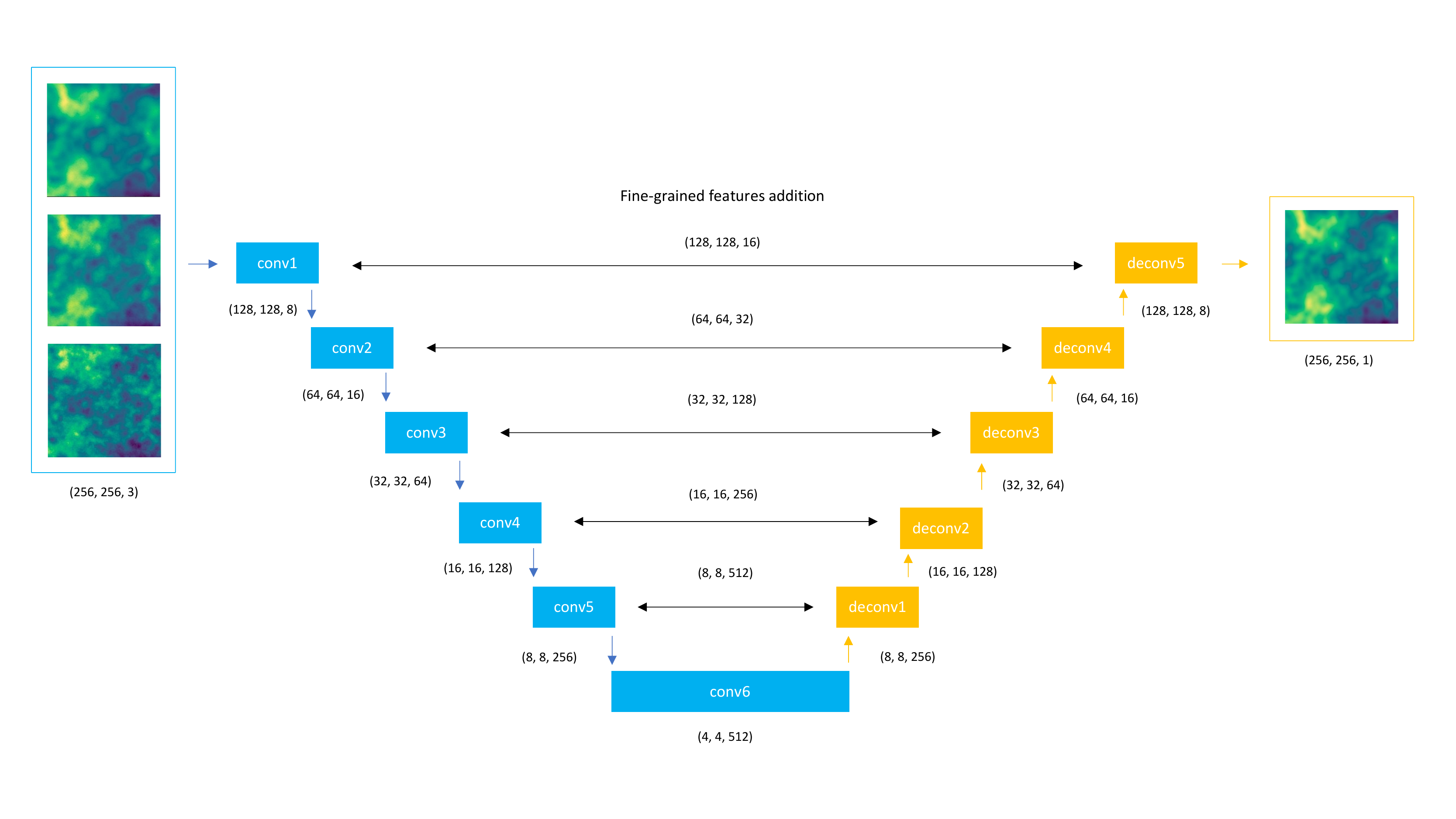}
\caption{Architecture of CENN. It has a convolutional block that produces 8 feature maps. After this, the space dimensionality increases to 512 feature maps through five more convolutional blocks. These layers are connected to deconvolutional blocks that decrease the space dimensionality to one feature map in the last deconvolutional block. Fine-grained features are added from each convolution to its corresponding deconvolution.}
\label{Fig 2.}
\end{figure*}

\subsection{Architecture}
\label{sec:FCNN}

Our topology is an FCN based on the U-Net architecture \citep{RON15}. The inputs of the network are three patches (256\hspace{1pt}$\times$\hspace{1pt}256 pixels) with the total sky emission, one for each frequency at the same position. The output is a single patch, with the same size as the input patches, containing the recovered CMB signal. The label for minimising the loss function is the true CMB signal on each simulation. The network was trained during 500 epochs with 60 000 simulations divided into minibatches of 32.

The architecture of CENN is shown in Figure~\ref{Fig 2.} and is detailed as follows: firstly, it has a set of six convolutional blocks, all of them formed by convolutional and pooling layers with 8, 2, 4, 2, 2, and 2 kernels of sizes of 9, 9, 7, 7, 5, and 3, respectively. The number of filters is 8, 16, 64, 128, 256, and 512, respectively. They have a subsampling factor of 2. The padding type "Same" is added in all the layers in order to add some space around the input data or the feature map, to deal with possible loss in width and/or height dimension in the feature maps after having applied the filters. The activation function is leaky ReLU in all the layers. Secondly, the FCN has a set of six deconvolutional blocks that are formed by deconvolutional and pooling layers with 2, 2, 2, 4, 2, and 8 kernels of sizes of 3, 5, 7, 7, 9, and 9, respectively. The number of filters is 256, 128, 64, 16, 8, and 1. Their subsampling factor is 2. The padding type "Same" is also added in all the layers, and the activation function is the leaky ReLU. 

In order to take fine-grained features in the image into account \citep{WEI21}, we have added layers to the architecture that connect the convolutional and deconvolutional blocks. This addition doubles the space of the feature maps before each deconvolutional block. The main goal for these layers is to help the network to predict low-level features with the deconvolutional blocks by taking high-level features into account that are inferred by the convolutional blocks. After each fine-grained layer, a deconvolutional block is connected in order to predict patterns that are dependent on the high and low-level features. In our model, the addition of these layers is related to the task of predicting small-scale regions of the CMB signal by taking already inferred large-scale structures into account. An example of CENN output patches is shown in the third column of Figure~\ref{Fig 1.}.

\section{Results}
\label{sec:results}

\begin{figure*}[ht]
\centering
\includegraphics[width=15cm]{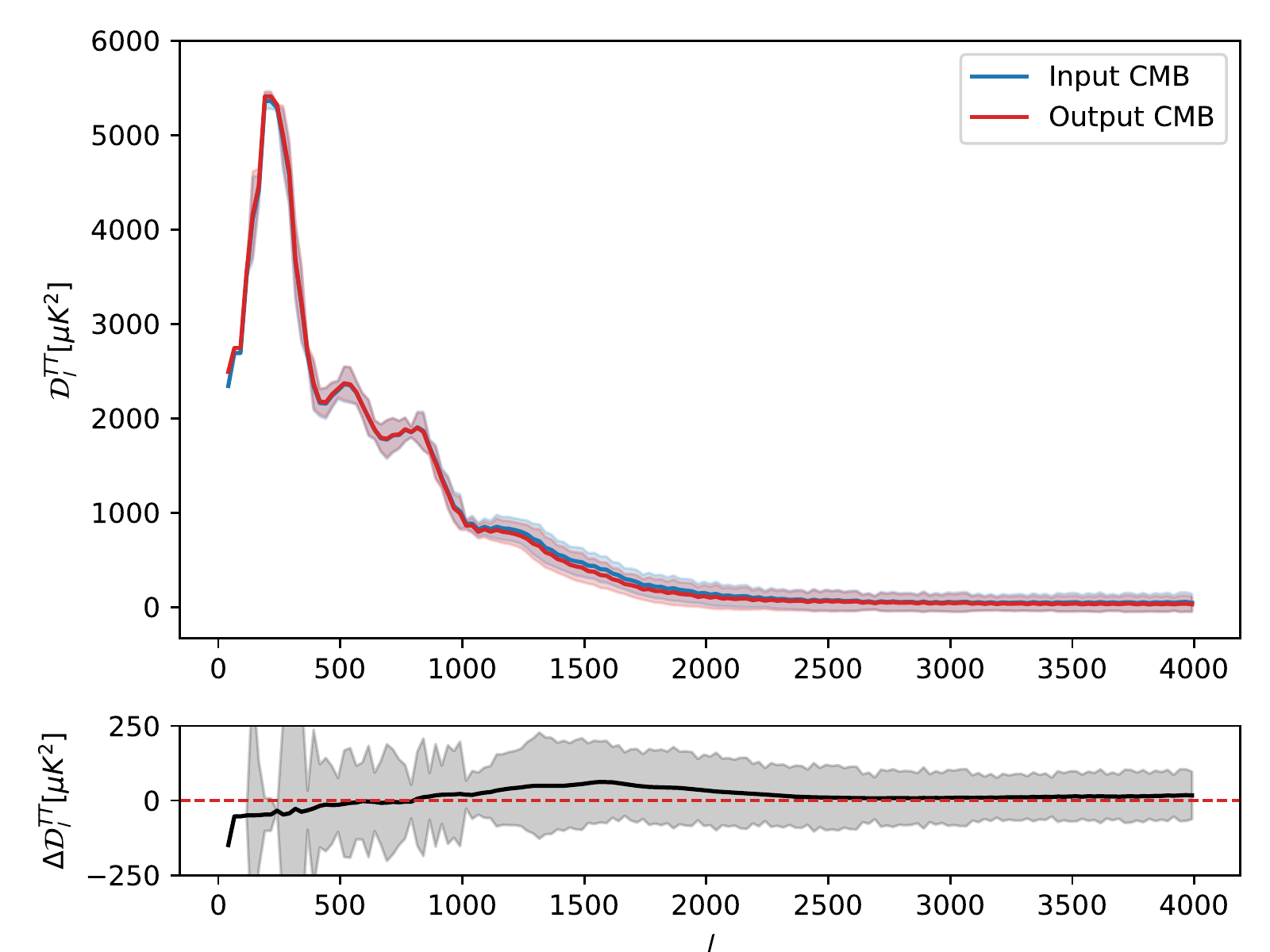}
\caption{Mean CMB power spectrum comparison computed over the entire validation dataset. The true CMB is represented as a blue line, and the output from CENN is shown as a red line. The corresponding uncertainties, computed as the standard deviation of each bin, are the blue and red areas, respectively. The difference between input and output is plotted in the bottom panel as a black line, and its uncertainty is the grey area.}
\label{Fig 3.}
\end{figure*}

After CENN was trained, we validated it using a dataset of 6~000 simulations of random patches in the whole sky and using three different validation datasets of 2 000 simulations each in different latitude ranges: the first dataset at $0^\circ<|b|<5^\circ$, the second dataset at $5^\circ<|b|<30^\circ$ , and the third dataset at $30^\circ<|b|<90^\circ$. All the validation datasets consisted of simulations that were different from the training dataset and, in particular, the CMB signal was from another PLA simulation. The analysis of the results is split into three parts: firstly, we study the CMB power spectrum in the whole sky in Sect. 4.1. Secondly, we analyse both the mean residuals and the residuals of the mean patch in the all sky case and in the three latitude intervals in Sect. 4.2. Thirdly, we estimate the levels of contamination at small scales in Sect. 4.3.

\subsection{Power spectra}
\label{sec:PowerSpectra}

We compare in the top panel of Fig.3 the power spectrum of the true CMB maps with the power spectrum from the outputs of CENN, validated over the whole sky. More precisely, we compare the average of the power spectra computed on the input (blue line) and output (red line) sky patches. The power spectrum was rebinned using a step of 50 up to $l$ = 1 000 and a step of 200 above $l$ > 1 000. The standard deviation of each bin for the input and output patches is also shown (blue and red areas, respectively), which is considered the uncertainty of each quantity. 
Furthermore, the bottom panel shows the difference between the input and the output average power spectra (black line) and its uncertainty (grey area). 

We recover the CMB power spectrum with a mean difference between input and output of 13 $\pm$ 113 $\mu K^{2}$ for all multipoles and -18 $\pm$ 150 $\mu K^{2}$ for multipoles up to $l \sim$ 1 000.  \citet{PET20} recovered the CMB with a difference of 3.8 $\mu K^{2}$ up to the same multipoles after masking map pixels with a predicted standard deviation of $>$50 $\mu K^{2}$. This was a per-pixel error comprising both statistical and model uncertainties. This was not the case in this work. However, in our model, we tested CENN without the use of any pixel mask. At the first multipoles, the CMB cannot be properly recovered because we work with patches of the sky. We therefore have poor statistics at the largest angular scales. Moreover, between $l \sim$ 1 000 and $l \sim $ 2 500, we obtain a CMB power spectrum with a difference of $20 \pm 100$ $\mu K^{2}$ with respect to the input data, while \citet{PET20} obtained a mean value above $100$ $\mu K^{2}$. This result is interesting mostly because the traditional component separation methods applied to \textit{Planck} data \citep{PLA_18_IV} also failed at these multipoles, and they needed to mask the Galactic region and the brightest PS to avoid this contamination. 
Our method can be used to recover the CMB in the entire sky, without applying any sky mask at all. This means that on the one hand, we successfully recover the CMB signal even in the region that is mostly affected by galactic emission and generally avoided by traditional component separation methods. On the other hand, because we do not need to mask the PS, we successfully recover the signal also at large angular scales. Therefore, it is quite likely that our network will recover the CMB signal with satisfying accuracy and reliability in future CMB experiments with better angular resolution than \textit{Planck}.

\subsection{Residuals}
\label{sec:ResidualMaps}

\begin{table*}[ht]
    \centering
    \caption{Mean value of the residual patch (top row) and its root mean-squared value (bottom row) for each sky latitude and for the whole sky.}
    \begin{tabular}{cccccccc}
    \hline\hline
    & & $0^\circ < b < 5^\circ$ & $5^\circ < b < 30^\circ$ & $30^\circ < b < 90^\circ$ & All sky\\
    \hline
    & Mean residual patch ($\mu K$) & 5.072  & - 0.026 & 5.238 & - 0.83 &  & \\
    \hline
    & RMS residual patch ($\mu K$) & 7.732 & 7.536 & 6.85 & 7.372 &  &  \\
    \hline\hline
    \end{tabular}
    \label{tab:datos}
\end{table*}

\begin{figure}[ht]
\centering
\includegraphics[width=9cm, height=6cm]{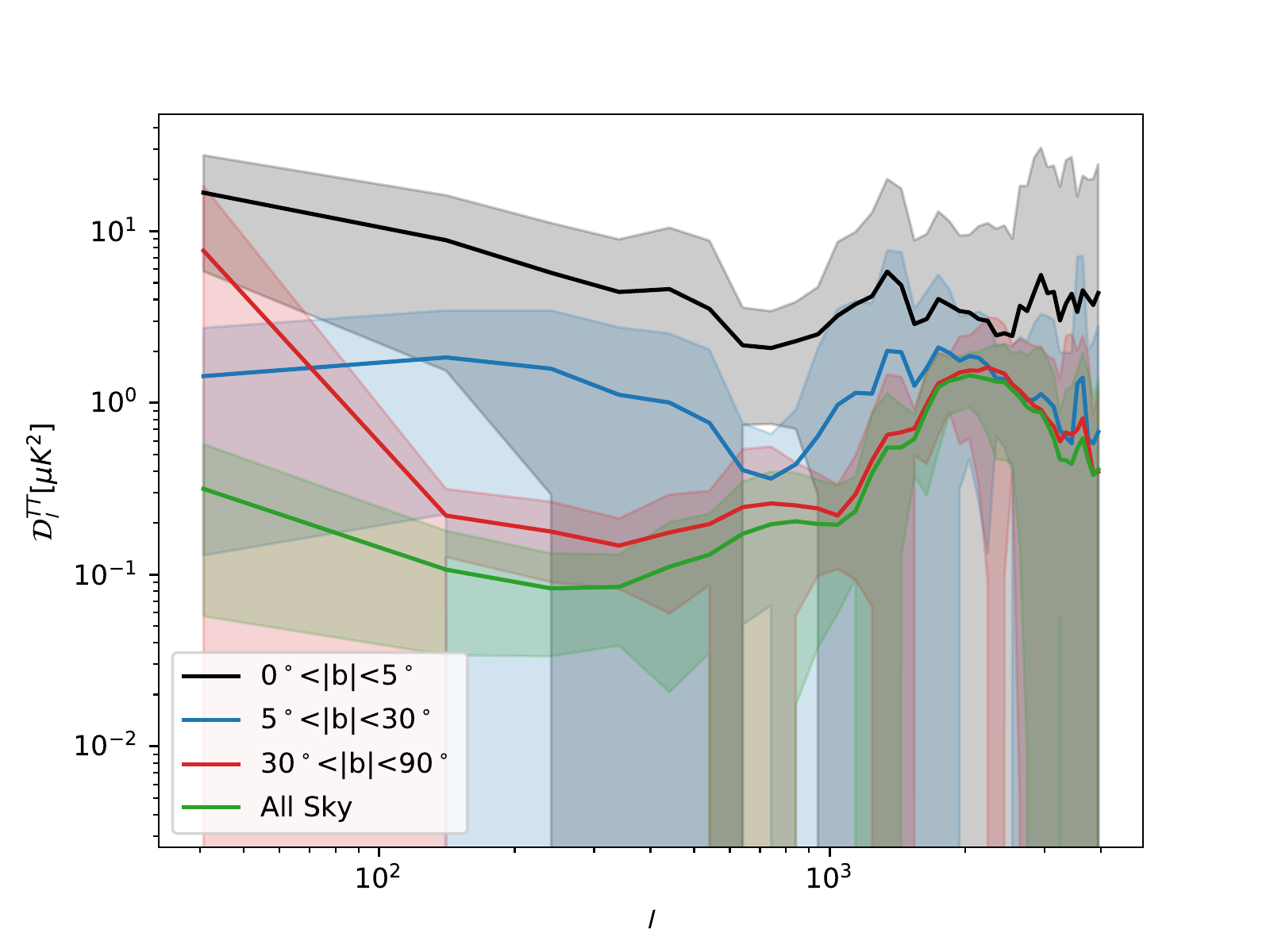}
\caption{Power spectrum of the mean residual patches for the whole sky (green continuous line) and for $0^\circ < |b| < 5^\circ$ (continuous black line),  $5^\circ < |b| < 30^\circ$ (continuous blue line),  $30^\circ < |b| < 90^\circ$ (continuous red line). The corresponding coloured areas are the standard deviation of each bin.}
\label{Fig 4.}
\end{figure}

\begin{figure*}[ht]
\centering
\includegraphics[width=17cm]{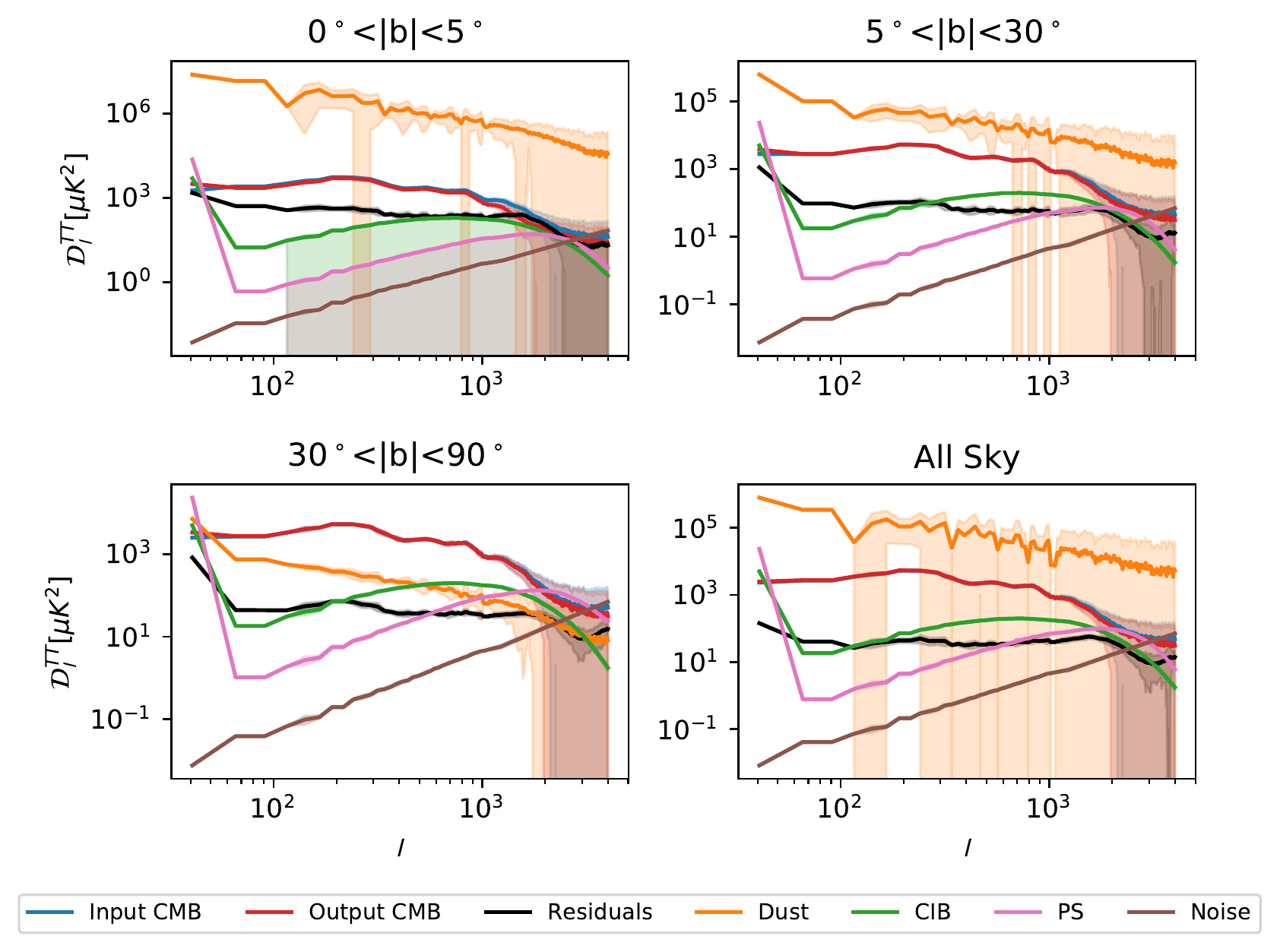}
\caption{Mean power spectrum of the residuals (black) and its uncertainty (grey area) for $0^\circ < |b| < 5^\circ$ (top left panel),  $5^\circ < |b| < 30^\circ$ (top right panel),  $30^\circ < |b| < 90^\circ$ (bottom left panel) and for the whole sky (bottom right panel) against the input and output CMB (blue and red lines, respectively) and their uncertainties (blue and red areas, respectively). In all cases, the uncertainties are the standard deviation of each bin. The contribution of each foreground in the simulations through their power spectra is also plotted. The thermal dust is represented by the orange line, the PS and CIB by the pink and green lines, respectively, and the instrumental noise is shown by the brown line.}
\label{Fig 5.}
\end{figure*}

In this section, we study the mean power spectrum of the residual maps (i.e. the difference between input CMB and CENN output patches) and the power spectrum of the mean residual patch. Firstly, we computed the mean and the root mean-squared (RMS) of the residual patch for the whole sky and for the three different latitude intervals. Table \ref{tab:datos} lists the mean value (top row) and the RMS value (bottom row) for $0^\circ<|b|<5^\circ$ (second column), $5^\circ<|b|<30^\circ$ (third column), $30^\circ<|b|<90^\circ$ (fourth column), and the whole sky (fifth column).
 
The mean value is 5.072 $\mu K$ for the inner Galactic region. This is close to the value for the extragalactic region, 5.238 $\mu K$. At intermediate latitudes, the mean value is -0.026 $\mu K$. Finally, when the whole sky is taken into account, we obtain a mean value of -0.83 $\mu K$. For the RMS, we obtain 7.732 $\mu K$ for the Galactic region and 7.536 $\mu K$ for the intermediate region, which is a lower number than was reported in \citet{LEA08} for some of the component separation methods used in \textit{Planck}. More precisely, they presented an RMS value of about 15 and 13 $\mu K$ for \texttt{SMICA} and 18 and 16 $\mu K$ for \texttt{SEVEM} for the latitude intervals, both obtained using realistic simulations.

In Figure \ref{Fig 4.} we compare the power spectrum of the mean residual patch, that is, firstly we computed the mean residual patch for each validation dataset, and after this, we estimated its power spectrum. This type of study is not only useful for a comparison against other segmentation methods, but also allow us to improve our statistical knowledge of the method, for example by identifying possible systematic biases such as artefacts near the borders, which are common for methods working with patches.

The residuals are lower than 15 $\mu K^{2}$ for multipoles up to $l \sim $ 4 000. In the inner Galactic region, the strong contamination implies 7 $\pm$ 25 $\mu K^{2}$, that is, only a higher residual for all multipoles with respect to the other cases. This is an important achievement because this region is usually avoided when more traditional component separation methods are used.

Although it is still a region with high Galactic contamination, our method performs well in the intermediate case, obtaining residuals of about 2 $\pm$ 10 $\mu K^{2}$ for multipoles up to $l \sim $ 4 000. As in the inner Galactic case, this region was usually partially masked by the component separation methods used in \textit{Planck}. Finally, in the extragalactic region, the power spectrum of the mean residual patch shows the expected behaviour: it has lower values overall because in this region, the Galactic contamination is significantly lower than in the other cases. 

The residual behaviour remains almost constant even at high multipoles. More precisely, for $l > $ 1 000, we obtain values of about 3 $\pm$ 11 $\mu K^{2}$ for $0^\circ<|b|<5^\circ$, 1 $\pm$ 2 $\mu K^{2}$ for $5^\circ<|b|<30^\circ$, 0.5 $\pm$ 0.6 $\mu K^{2}$ for $30^\circ<|b|<90^\circ$ , and 0.8 $\pm$ 1.5 $\mu K^{2}$ for the whole sky, which are still a low values for the residuals. This suggests that in principle, masking might not be needed with our network. This gives us the advantage that we do not have to deal with the holes left by PS masks in CMB reconstructed maps. The performance at these multipoles is probably due to the fact that the network is trained to segmentate diffuse signals, which have a completely different characteristics than the point-like signals. Then, the PS contamination is relatively irrelevant while reconstructing the CMB signal with CENN. 

In Figure \ref{Fig 5.} we compare the mean power spectrum (black line) of each residual patch in the validation datasets with the input and recovered CMB signal and the level of contamination of each foreground in the simulations (the thermal Sunyaev-Zel'dovich effect appears in the 143 and 353 GHz input simulations during validation, but because the output is at 217 GHz, the contribution is neglected). The corresponding uncertainties for each quantity (i.e. the standard deviation, as in the above subsection) are represented as coloured areas.

The residuals depend strongly on contamination levels. In the Galactic plane region, we obtain residuals of about 700 $\pm$ 60 $\mu K^{2}$. In the other Galactic region, the residuals decrease to 80 $\pm$ 30 $\mu K^{2}$ . However, the contamination levels in these regions are about three and two orders of magnitude higher, respectively. In the extragalactic region, the impact of contaminants is not as high as in the other cases, and this is also true for the residuals, which are above 30 $\pm$ 20 $\mu K^{2}$. Finally, with random simulations in the whole sky, we obtain residuals of about 20 $\pm$ 10 $\mu K^{2}$, although the contamination levels are almost two orders of magnitude.

At small scales, the residuals show a slight bump in all cases, which is also visible in Figure \ref{Fig 4.}. It is due mainly to extragalactic foregrounds. However, it should be noted that for multipoles $l > 2 500$, where the \textit{Planck} beam caused the foregrounds to decrease, the residuals remain constant and at lower levels than at large scales. The continuous brown line in Fig. 5 shows that the instrumental noise is strong in this region, however. This is the first indication that these types of ML models might be able to handle instrumental effects. This behaviour of CENN should be studied in detail and tested in future works because it is crucial to constrain both instrumental and systematics effects for primordial B mode detection.

\subsection{Contamination at small scales}
\label{sec:SmallScales}

\begin{figure*}[ht]
\centering
\includegraphics[width=15.5cm, height=3.5cm]{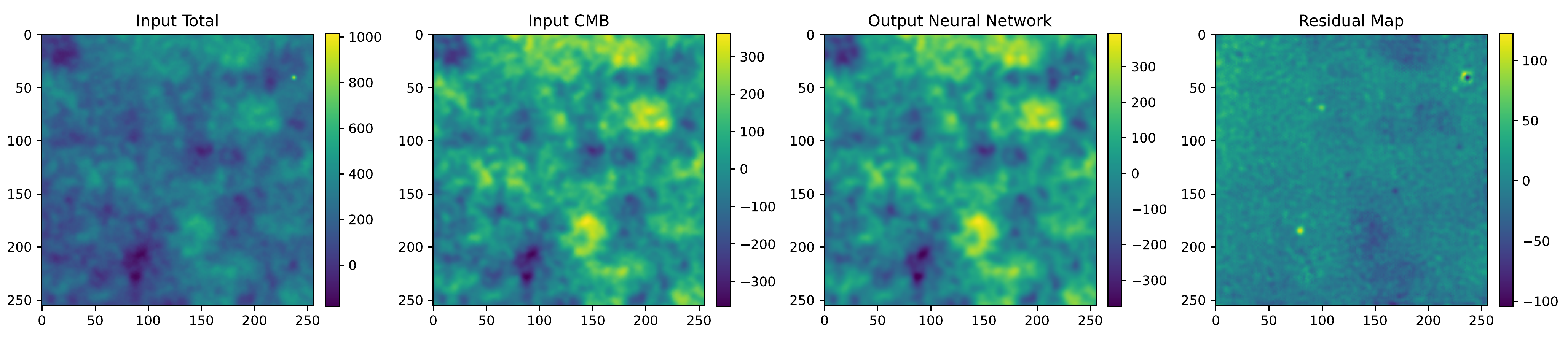}
\includegraphics[width=12cm]{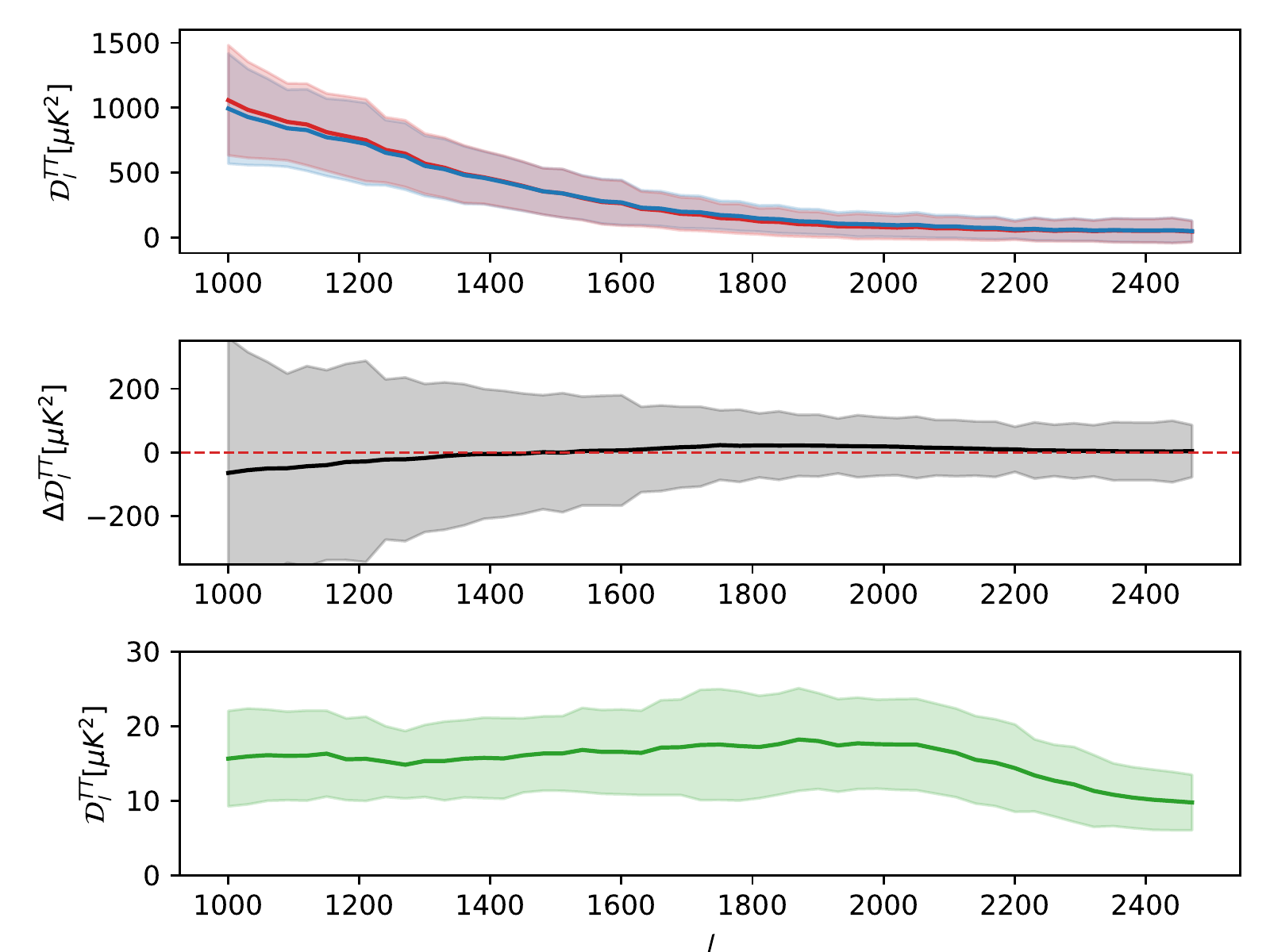}
\caption{Analysis of a patch highly contaminated at small scales. Top panel: Sky patch selected with high contamination at small scales. The patches represent from left to right all the emission, the CMB label, the neural network output, and the residual map computed as the difference between the second and third columns. The frequency for all patches is 217 GHz, and their temperature (in $\mu K_{CMB}$) is shown in the right vertical bars. Bottom panel: CMB power spectrum comparison computed over the patch shown in the top panel. The true CMB is represented in the top panel as a blue line, and the output from CENN is shown as a red line. The corresponding uncertainties (i.e. the standard deviation) are shown as blue and red areas, respectively. The difference ($\Delta \mathcal{D}_{l}^{TT}$) between input and output is plotted in the middle panel as a black line, and its uncertainty is the grey area. The power spectrum of the residual map is represented as a green line in the bottom panel. The green area is its uncertainty (also the standard deviation).}
\label{Fig 6.}
\end{figure*}

In this subsection we focus on studying the performance of the network against contamination at small scales by analysing a specific patch that was selected for its much higher visual contamination at small scales with respect to the average case. The patch is represented in the top panel in Figure \ref{Fig 6.} (units in $\mu K$): from left to right, the total patch with all the contaminants and the CMB added together, the input CMB for the same patch, the network output, and the residual (the difference between the input CMB and the output) patches.

The small-scale structure can be seen in the pixels [237, 41], [100, 70], [169, 148], and [80, 185], mostly in the residual patch. Despite the high contamination, the residuals are relatively low. The mean value is -0.72$\pm$15.52 $\mu K$. The highest and lowest values that can be also seen by eye in the residual patch correspond to point-like objects that contaminate the network output at small scales. In particular, the pixels at [237, 41] are the brightest in the input total patch with a value of about 1 000 $\mu K$. This point-like structure is still present in the output patch with a value lower than 300 $\mu K$ and is clearly visible in the residual patch with absolute values of about 100 $\mu K$. These numbers suggests that the network still performs well even for this problematic patch because the contamination from small scales is notably reduced.

In the bottom panel in Figure \ref{Fig 6.} we also compare the power spectra of the input, output, and residual patches of this sky area. The rebinning has a step of 50 until $l$ < 1 000, and a step of 200 above $l$ = 1 000.
In the top panel we represent the input and output CMB power spectra as blue and red lines, respectively, with their uncertainties, estimated as the standard deviation, plotted as blue and red areas, respectively. 
In the middle panel, we show the difference between input and output power spectra. The grey area is the uncertainty. In the bottom panel, we plot the power spectrum of the residual patch, and the green area is the uncertainty.

CENN recovers the CMB power spectrum with a mean difference between 0 and 30 $\mu K^{2}$ (black line) for the entire range of multipoles. In particular, between $l$ = 1 200 and $l$ = 1 800, the difference between input and output is higher, with a value of about 50 $\mu K^{2}$. The estimated power spectrum of the residuals (green line) at the same multipoles has 40 $\mu K^{2}$, decreasing at higher multipoles. 

When our network is applied without any mask, the residual PS slightly affects the network output, but mainly at multipoles around 2 000. The residuals at higher multipoles are negligible, and no residual contamination is related to the instrumental noise, at least at the \textit{Planck} noise level. Moreover, we tested the performance of the network in patches with five and ten times higher instrumental noise than the \textit{Planck} noise, finding that the performance at high multipoles is similar in the three cases, and the error is almost negligible. This behaviour is very different from that of the traditional component separation methods, for which above a certain multipole, $l >$ 2 500, the power spectrum is completely dominated by the PS and instrumental noise residuals, even after a masking procedure. 
\section{Conclusions}
\label{sec:conclusions}

We developed a new component separation method based on artificial neural networks for future CMB experiments. More precisely, we trained a fully convolutional neural network called CENN with realistic simulations of the central channels of the \textit{Planck} mission to extract the CMB signal in total intensity from the other foregrounds, which are thermal dust from our Galaxy, the CIB signal, the thermal Sunyaev-Zel'dovich effect, the contribution from PS (radio and IR late-type galaxies), and instrumental noise. The frequencies are the 143, 217, and 353 GHz channels.

To train CENN, we used 60 000 realistic simulations. Each simulation had three image patches with an area of 256\hspace{1pt}$\times$\hspace{1pt}256 pixels containing all the emissions and an additional map with only the CMB contribution at 217 GHz, which was used to learn this signal by the neural network via the optimiser and the loss function. In addition to this dataset, we used 6 000 simulations to test the network during the training. After the training, we validated the network with different simulations, which had patches of the CMB signal extracted from a different PLA simulation than the one used for training the network. We developed four different validation datasets: one formed by 6 000 random simulations at all sky, and another three with 2 000 simulations each set at $0^\circ<|b|<5^\circ$, $5^\circ<|b|<30^\circ$ , and $30^\circ<|b|<90^\circ$ degrees of latitude. After validating CENN, we analysed the results by rebining the data and computing the mean power spectra of the input and output CMB from the neural network, and the difference between them, in all cases considering the standard deviation as the uncertainty of each bin. We obtain a mean difference of 13 $\pm$ 113 $\mu K^{2}$ for multipoles up to above 4 000.

We computed the mean power spectrum of the residuals (i.e. the difference between input and output CMB), obtaining 700 $\pm$ 60 $\mu K^{2}$ for  $0^\circ<|b|<5^\circ$, 80 $\pm$ 30 $\mu K^{2}$ for $5^\circ<|b|<30^\circ$ , and 30 $\pm$ 20 $\mu K^{2}$ for $30^\circ<|b|<90^\circ$. We also analysed the mean and standard deviation patches for each latitude and for the whole sky. We obtain a RMS value of 7.732 $\mu K$ for $0^\circ<|b|<5^\circ$, 7.536 $\mu K$ for $5^\circ<|b|< 30^\circ$, 6.85 $\mu K$ for $30^\circ<|b|< 90^\circ$ , and 7.372 $\mu K$ for all sky. Furthermore, we computed their power spectra, obtaining 7 $\pm$ 25 $\mu K^{2}$ for $0^\circ<|b|<5^\circ$, 2 $\pm$ 10 $\mu K^{2}$ for $5^\circ<|b|<30^\circ$, and 2 $\pm$ 3 $\mu K^{2}$ for $30^\circ<|b|<90^\circ$. 

Finally, we studied the performance of CENN in recovering the CMB signal in a patch with high contamination at small scales. We obtain a value of 50 $\pm$ 250 $\mu K^{2}$ for the difference between input and output CMB power spectra, and residuals of about 40 $\pm$ 10 $\mu K^{2}$ at multipoles between $l$ = 1 200 and $l$ = 1~800.

These results show that our model is reliable not only at the multipoles analysed in \textit{Planck} ($l$ <= 2 500), but also for higher multipoles. Therefore, it seems to be a promising model for extracting the CMB signal for future experiments with higher angular resolution than \textit{Planck}.

Moreover, it has other advantages: the first advantage is that our model is evaluated in the whole sky without the use of any confidence mask to avoid strong Galactic contamination regions. Therefore, a more realistic CMB map might be obtained with respect to the inpainted version at the Galactic plane used in \textit{Planck}. The second advantage is that the PS or galaxy cluster contamination is very small and affects the output power spectrum only in a short multipole range around $l \sim$ 2 000. These results can easily be upgraded using a dedicated masking scheme or providing information about the PS and galaxy clusters during the training. These potential upgrades are beyond the scope of the current paper and will be considered for the application of the network to real data. 

Both conclusions about the reliability of CENN with the foregrounds at large and small scales might be a first indication that the network might be able to separate the CMB from the foregrounds by learning their non-Gaussian structures \citep{COU19}, which could be an advantage of this approach with respect to the traditional approaches. Efforts to study the network performance with these non-Gaussian structures will be crucial in order to detect primordial B-modes on polarisation data. However, this is beyond the scope of this work and will be considered in future developments. Moreover, our results suggest that CENN might be able to perform a component separation well even in single-frequency maps, which is worth to be tested in the future as well. Lastly, when it is trained, our model can extract the CMB in microwave maps almost immediately, which is fundamental for future satellites with high amounts of data.

As we explained in Section \ref{sec:simulations}, we have trained CENN with realistic simulations from the Planck Sky Model. However, these simulations do not not take the non-Gaussian behaviours of the foregrounds of the microwave sky properly into account, especially at the frequencies analysed in this work, where thermal dust is the most significant foreground and it is known to have a strong non-Gaussian behaviour. Therefore, our results could be biased by this approximation of the microwave sky when validating it with real data, even if, as commented above, our network seems to rely on such non-Gaussian structures of the different emissions to recover the CMB signal.
Because this issue is one of the major limitations of this kind of ML models, the network performance in total intensity and polarisation must be further tested in the future by training it with more realistic simulations, for example the outputs by ForSE (the generative adversarial neural network developed by \citet{KRA21}), which can include small-scale non-Gaussian features extended to 12 arcmin.

\begin{acknowledgements}
We warmly thank the anonymous referee for the very useful and constructive comments on the original manuscript. JMC, LB, JGN, MMC and DC acknowledge financial support from the PGC 2018 project PGC2018-101948-B-I00 (MICINN, FEDER). JMC also acknowledges financial support from the SV-PA-21-AYUD/2021/51301 project. MMC also acknowledges to be granted by PAPI-20-PF-23 and PAPI-21-PF-04 (Universidad de Oviedo). CB acknowledges support from the ASI COSMOS and LiteBIRD Networks (cosmosnet.it), as well as by the INFN INDARK and LiteBIRD grants. JDS, MLS and JDCJ acknowledge financial support from the I+D 2017 project AYA2017-89121-P and support from the European Union’s Horizon 2020 research and innovation programme under the H2020-INFRAIA-2018-2020 grant agreement No 210489629.\\
This research has made use of the python packages \texttt{matplotlib} \citep{matplotlib}, \texttt{Keras} \citep{KER}, and \texttt{Numpy} \citep{numpy}, also the \texttt{HEALPix} \citep{GOR05} and \texttt{healpy} \citep{zon19} packages.
\end{acknowledgements}

%
%
\bibliographystyle{aa}
\bibliography{SDNN}


\appendix

\section{Learning process of CENN}
\label{sec:appendix}

Kernels and filters cited in section \ref{sec:Model} are adjusted in each epoch by updating non-linear parameters $\theta_{t + 1}$ as 
\begin{equation}
    \theta_{t + 1} \hspace{1pt} = \hspace{1pt} \theta_{t} \hspace{1pt} + \hspace{1pt} \Delta \hspace{1pt} \theta_{t},
\label{eq:parameter}
\end{equation}
where $\theta_{t}$ is the parameter of the previous epoch, and 
\begin{equation}
    \Delta \theta_{t} \hspace{1pt} = \hspace{1pt} - \hspace{1pt} \alpha \hspace{1pt} g_{t},
\label{eq:parameter_variation}
\end{equation}
where $g_{t}$ is the gradient value at the epoch $t$. Furthermore, $\alpha$  depends on the selected optimisation algorithm. Because our training was performed using the adaptive gradient algorithm \citep[AdaGrad,][]{Duchi2011}, the parameter $\alpha$ in Eq. \eqref{eq:parameter_variation} varies as
\begin{equation}
    \alpha \hspace{1pt} = \hspace{1pt} \frac{\eta}{\sqrt{G_{t} \hspace{1pt} + \hspace{1pt} \epsilon}},
\label{eq:adagrad}
\end{equation}
where $G_{t}$ is a diagonal matrix at epoch $t,$ for which the diagonal elements correspond to the sum of the squares of the past gradients, $\epsilon$ is a smoothing term that avoids division by zero, and $\eta$ is the learning rate, with a value of 0.05 in our model. 

When all minibatches of the training dataset flow through all the layers, the gradient in equation \eqref{eq:parameter_variation} is then computed in the last deconvolutional block by taking the derivative of the mean-squared error loss function (\textit{E}) of equation \eqref{eq:mse} over the output patch $O_{\nu_{0}}$, where the subindex $\nu_{0}$ represents the frequency channel of the output patch (217 GHz in this work). Therefore, the gradient at the last deconvolutional block is easily estimated by using
\begin{equation}
    g_{t} \hspace{1pt} = \hspace{1pt} \frac{\partial \hspace{1pt} E}{\partial \hspace{1pt} O_{\nu_{0}}} \hspace{1pt} = \hspace{1pt} \frac{\partial}{\partial \hspace{1pt} O_{\nu_{0}}} \left( \frac{1}{2} \hspace{1pt} | \hspace{1pt} O_{\nu_{0}} \hspace{1pt} - \hspace{1pt} l_{\nu_{0}} \hspace{1pt} | \hspace{1pt}^{2} \right) \hspace{1pt} = \hspace{1pt} | \hspace{1pt} O_{\nu_{0}} \hspace{1pt} - \hspace{1pt} l_{\nu_{0}} \hspace{1pt} |,
\label{eq:last_deconvolutional}
\end{equation}
where $l_{\nu_{0}}$ is the true value (also known as label or target), that is, the signal we wish to extract from the maps, which in our case is the CMB. For the previous deconvolutional block, the model uses this value, that is, the gradient of the subsequent block, and the chain rule to update its weights and filters. Therefore, for the previous deconvolutional block the gradients are
\begin{equation}
\begin{split}
    g_{t, \hspace{1pt} W} \hspace{1pt} = \hspace{1pt} \frac{\partial \hspace{1pt} O_{\nu_{0}, \hspace{1pt} LD}}{\partial \hspace{1pt} W_{LD - 1}} \hspace{1pt} & = \hspace{1pt} \frac{\partial \hspace{1pt} O_{\nu_{0}, \hspace{1pt} LD}}{\partial \hspace{1pt} O_{LD - 1}} \frac{\partial \hspace{1pt} O_{LD - 1}}{\partial \hspace{1pt} W_{LD - 1}} \hspace{1pt} = \\
    & \hspace{1pt} = \hspace{1pt} g_{t, \hspace{1pt} LD} \hspace{1pt} \frac{\partial \hspace{1pt} O_{LD - 1}}{\partial \hspace{1pt} W_{LD - 1}} \hspace{1pt} = \hspace{1pt} g_{t, \hspace{1pt} LD} \hspace{1pt} f_{LD - 1} 
\end{split}
\label{eq:gradients1}
\end{equation}
\begin{equation}
\begin{split}
    g_{t, \hspace{1pt} f} \hspace{1pt} = \hspace{1pt} \frac{\partial \hspace{1pt} O_{\nu_{0}, \hspace{1pt} LD}}{\partial \hspace{1pt} f_{LD - 1}} \hspace{1pt} & = \hspace{1pt} \frac{\partial \hspace{1pt} O_{\nu_{0}, \hspace{1pt} LD}}{\partial \hspace{1pt} O_{LD - 1}} \frac{\partial \hspace{1pt} O_{LD - 1}}{\partial \hspace{1pt} f_{LD - 1}} \hspace{1pt} = \\
    & = g_{t, \hspace{1pt} LD} \frac{\partial \hspace{1pt} O_{LD - 1}}{\partial \hspace{1pt} f_{LD - 1}} \hspace{1pt} = \hspace{1pt} g_{t, \hspace{1pt} LD} \hspace{1pt} W_{LD - 1}, 
\end{split}
\label{eq:gradients2}
\end{equation}
where $g_{t, \hspace{1pt} W}$ and $g_{t, \hspace{1pt} f}$ are the gradients used to update the weights and the filters, respectively. The final derivatives in equations \eqref{eq:gradients1} and \eqref{eq:gradients2} are computed using the fact that a matrix convolution operation is defined as
\begin{equation}
    O_{i, \hspace{1pt} j} = \sum_{m = 0}^{M} \hspace{1pt} \sum_{n = 0}^{N} \hspace{1pt} f(i - m, j - n) \hspace{1pt} W(m, n). 
\label{eq:convolution_matrix}
\end{equation}
Moreover, the achronyms LD and LD-1 correspond to the last deconvolutional and its previous blocks, respectively. The remaining deconvolutional blocks have similar expressions. The convolutional blocks have the same derivation as Eqs. \eqref{eq:gradients1} and \eqref{eq:gradients2} to update their corresponding weights and filters, respectively. More precisely, for the last convolutional block,
\begin{equation}
    g_{t, \hspace{1pt} W} \hspace{1pt} = \hspace{1pt} g_{t, \hspace{1pt} FD} \hspace{1pt} f_{LC} 
\label{eq:gradients_convolutional1}
\end{equation}
\begin{equation}
    g_{t, \hspace{1pt} f} \hspace{1pt} = \hspace{1pt} g_{t, \hspace{1pt} FD} \hspace{1pt} W_{LC}, 
\label{eq:gradients_convolutional2}
\end{equation}
where the achronyms FD and LC correspond to the first deconvolutional and the last convolutional block, respectively. For the remaining convolutional blocks, both filters and weights are updated following the same way as in Eqs. \eqref{eq:gradients_convolutional1} and \eqref{eq:gradients_convolutional2}.

\end{document}